\documentclass[11pt]{article}
\usepackage{amsmath,amssymb}
\usepackage{graphicx}
\usepackage{cite}
\usepackage[labelfont=bf]{caption}
\title{A New Approach to Flatness, Horizon and Late-time Accelerating Expansion Problems on the basis of Mach's Principle}
\date{\today}
\begin{document}

\author{Onder Dunya\footnote{onder.dunya@boun.edu.tr}\,  and Metin Arik\footnote{metin.arik@boun.edu.tr}\\
\textit{Department of Physics, Bogazici University, Bebek, Istanbul, Turkey}}
\maketitle

\begin{abstract}

Based on the idea that the components of a cosmological metric may be determined by the total gravitational potential of the universe,  the scalar field $\phi=1/G$ in the Jordan-Brans-Dicke (JBD) theory is introduced as evolving with the inverse square of the scale factor.  Since the gravitational potential is related to the field $\phi$ resulting from Mach's principle and depends on time due to the expansion of space,  the temporal evolution of the field should be in accord with the evolution of time and space intervals in the metric tensor.  For the same reason,  the time dependence of the field makes these comoving intervals relative for different points on the time axis. Thus,  it is shown that introduction of the cosmic gravitational potential as a time dependent scalar field proportional to $1/a^2$ may resolve the flatness,  the horizon and the late-time accelerating expansion problems of the standard model of cosmology.  The luminosity distance vs redshift data of Type Ia supernovae is in agreement with this approach. \\

\end{abstract}

\section{Introduction}
The standard model of cosmology is based upon the Einstein field equation that can be derived from the Einstein-Hilbert action and the Friedmann-Lemaitre-Robertson-Walker (FLRW) metric that represents a space-time in which the space expands with the scale factor and the time is absolute for all observers.  While the right hand side of the field equation is designated by the energy-momentum tensor $T_{\mu\nu}$,  the left hand side is determined by the geometry of space-time or simply by the Einstein tensor $G_{\mu\nu}$
\begin{equation}
G_{\mu\nu}=R_{\mu\nu}-\frac{1}{2}Rg_{\mu\nu}=8\pi G T_{\mu\nu},
\end{equation}
where $G$ is Newton's gravitational constant.
Furthermore,  the most general form of the FLRW metric without any perturbation,  which is used to draw a basic picture of the standard cosmology,  is written as
\begin{equation}
ds^2=-dt^2+a^2\left(\frac{dr^2}{1-\kappa r^2}+r^2d\Omega^2\right) .
\end{equation}
In Eq.(2),  $\kappa$ corresponds to the curvature parameter of space, which has the dimension of $length^{-2}$ ($c=1$ and subscript 0 denotes today's value of any cosmological parameter unless it is stated explicitly throughout the manuscript). Furthermore,  $\kappa$ can be negative, zero and positive for an open,  flat and closed universe respectively.  By using the FLRW metric for the field equation, one obtains the Friedmann equation.
\begin{equation}
H^2=\left(\frac{\dot{a}}{a}\right)^2=\frac{8\pi G}{3}\rho-\frac{\kappa}{a^2}
\end{equation}
Here,  $H$ is the Hubble parameter whose current value is $H_0=(67.4\pm 0.5)$ km/s/Mpc according to the latest observation\cite{2018planck} on the basis of the $\Lambda$CDM cosmology.  Another fundamental equation,  the continuity equation,  is derived from the conservation of energy-momentum tensor $\nabla_{\mu}T^{\mu\nu}=0$ as
\begin{equation}
\dot{\rho}+3\frac{\dot{a}}{a}(\rho+p)=0
\end{equation}
with the solution $\rho(t)=\rho_0 a^{-3(1+w)}$ for an equation of state $p=w\rho$.  Based on this solution,  while the matter density of the universe $\rho_m\sim \frac{1}{a^3}$,  the radiation density $\rho_r\sim \frac{1}{a^4}$ in which extra $1/a$ factor comes from the red-shift of photons due to the expansion of space.  Photons arriving us today from distant stars and galaxies,  are red-shifted by
\begin{equation}
z=\frac{\lambda_{ob}-\lambda_{em}}{\lambda_{em}},
\end{equation}
and the relation between the scale factor and the red-shift is
\begin{equation}
a(t)=\frac{1}{1+z} .
\end{equation}

By measuring the red-shift of photons and the luminosity distance of standard candles like Type Ia supernovae,  it is possible to determine how fast the expansion of the space is,  and to make inferences about essential quantities such as the age and the energy content of the universe.  In the standard model,  the luminosity distance\cite{carroll} is given by
\begin{equation}
d_{L}(z)=(1+z) \frac{H_{0}^{-1}}{\sqrt{|\Omega_{\kappa 0}|}}S_k\left(\sqrt{|\Omega_{\kappa 0}|} \int \frac{dz^{\prime}}{E(z^{\prime})}\right),
\end{equation}
where $\Omega_{\kappa 0}$ is today's value of the curvature density parameter,  $S_{\kappa}$ is a function for different spatial curvatures ($S_{\kappa}(\chi)$ is $\sin(\chi)$, $\chi$ and $\sinh(\chi)$ for a closed,  flat and open universe respectively) and for all types of energy contribution $E(z^{\prime})=\left[\Omega_{m 0}(1+z^{\prime})^3+\Omega_{r 0}(1+z^{\prime})^4+\Omega_{\kappa 0}(1+z^{\prime})^2\right]^{1/2}$.

In the last two decades, it has been unveiled that the expansion rate predicted by the standard model does not fit in the data of Type Ia supernovae\cite{1999perlmutter, 1998riess} when only the contributions of matter and radiation are taken into account.  According to the data the expansion is accelerating and the problem can be solved by adding some exotic constituents as a dark energy term.  The cosmological constant $\Lambda$, one of the candidates, has been introduced into the action.  However,  this type of solution brought its own problems such as the cosmic coincidence\cite{cosmiccoincidence} and the cosmological constant\cite{weinberg-cosmologicalconstant, martin2012} problems.  In the former one,  with "coincidence", it is meant that the era in which the cosmological constant dominates the evolution of the universe matches up with the period of time we started to make observations.  Second,  the cosmological constant is many orders of magnitude smaller than the value estimated in particle physics.  Even though some approaches\cite{weinberg-cosmologicalconstant, nobbenhuis} have been suggested,  they are found inconclusive.  So,  the problems still remain unsolved.

To explain the late-time accelerating expansion,  alternative gravity theories like scalar-tensor theories in which a scalar field is non-minimally coupled to the geometry of space-time or to matter fields in the action\cite{joyce2016} (depending on the frame chosen,  the Jordan frame or the Einstein frame respectively) have been proposed.  In the Jordan-Brans-Dicke (JBD) theory\cite{jordan, bransdicke},  which is one of the favored scalar-tensor theories,  a scalar field has been introduced with a motivation based on Mach's principle stating that inertial forces on a body are originating from the gravitational effects of matter distribution in the universe. 

The following relation, where the only time dependent parameter is the radius of the observable universe $R$ (this is a finite distance regardless of the spatial curvature) contains a connection between the gravitational potential of the universe and natural constants $G$ and $c$.
\begin{equation}
\frac{GM}{Rc^2}\sim 1
\end{equation}
Here,  $M$ is the finite mass of the observable universe.  According to Brans and Dicke,  this relation is not a coincidence and they argued that there should be at least one other time dependent parameter so that Eq.(8) has an explanation.  In their opinion, the most suitable one is the gravitational constant $G$,  and the relation (by taking $c=1$ again) is
\begin{equation}
\frac{1}{G}=\phi\sim \frac{M}{R} .
\end{equation}

Since the JBD theory was first introduced in 1961, it has been studied for many different cases in numerous ways. Considering that it tried to fully integrate Mach's principle into general relativity in the years it was first put forward,  it can be seen as a fundamental correction to the existing theory.  Over the years, it has become one of the modified gravity theories applied to explain the cosmic puzzles, in particular, the accelerating expansion in the late times of the universe.  In general, the solution requires a potential term of self-interacting JBD scalar field to be added in the action.  Thus, different extensions of the theory have emerged\cite{bertolami, sen2001late, senseshadri, sharif2012cosmic, bisabr2012cosmic, chakrabarti2022screening}.  One of the common points of such approaches is that the JBD coupling parameter should be negative in order to fit the data.  Although it has been deduced from solar system observations that this parameter must be a positive large number, its value at cosmic scales does not have to be necessarily the same.  In other words, it is a viable scenario for the JBD coupling parameter to be scale dependent.  Since the expansion of space is ignored in local observations, the solar system is a system that has reached equilibrium in this sense. However, this is not the case at cosmic scales.  

The late-time accelerating expansion of the universe is not the only problem with the standard model of cosmology.  The flatness and horizon problems are almost equally important and the theory of inflation\cite{guth, linde} is referred as a solution to these two.  Although the standard inflation theory has its own problems \cite{turok2002critical, brandenberger2000inflationary, martin2001trans, brandenberger2007conceptual, brandenberger2013trans}, it is also necessary to explain the formation of the Large Scale Structures (LSS) and the anisotropy in the Cosmic Microwave Background (CMB) in the $\Lambda$CDM cosmology which is the current concordance model.  Therefore,  it is also possible to see the applications of the JBD theory as the extensions of the standard theory of the inflationary expansion\cite{la1989extended, mathiazhagan1984inflationary, weinberg1989some, la1989prescription, linde1994hybrid, tahmasebzadeh2016brans, baptista1996density, kolb1990origin} in the early universe.  Although such approaches solve the problems of inflation theory up to a point, the search for a model that solves all three problems  and includes a specific JBD scalar field that evolves consistently between the two periods continues (because the horizon and flatness problems are related to the early universe,  while the accelerated expansion concerns the late times). The constraints, which have been put on the evolution of the JBD scalar field and the JBD coupling parameter by fitting the standard JBD theory and its extensions with various potential terms to the Type Ia supernova,  the CMB,  the Baryon Acoustic Oscillation (BAO) and other cosmological data sets (or the combinations of these),  can be found in the literature\cite{acquaviva2007observational, acquaviva2005structure, chen1999cosmic, wu2010cosmic, li2013constraints, ballardini2019testing, joudaki2022testing}.  However, since the model we come up with is a distinctive modification of the JBD theory rather than just adding a potential term and the observational tests mentioned above are model dependent, one should not question the validity of our model on the basis of these tests.

The aim of this article is to solve three basic problems at once,  by modifying the JBD theory without using any potential term and without including the inflationary expansion.  Our approach starts with the same relation in Eq.(8) but what we're trying to do is quite different from the JBD theory extensions aforementioned.  In addition,  since there will be no need for the dark energy and the inflation in the early universe,  our model is free from their problems. This is a more fundamental modification with a relation assumed to exist between the JBD scalar field and the metric tensor.  Although we will use the standard JBD action in this new model, the field equation will be different due to the assumption we make based on some motivations.  In the next section,  the FLRW metric will be coordinate transformed for a reason explained in detail.  In the third section,  the field equation of the modified JBD theory will be obtained.  Then, in the fourth section,  our main focus will be presenting possible solutions to the basic problems of the standard model of cosmology thanks to this modification.

\section{Transformed FLRW Metric}

As a starting point, we consider the perturbed metric due to the presence of a gravitational field in a non-expanding space. Its the most general form is given by
\begin{equation}
ds^2=-(1+2\Psi)dt^2+(1+2\Phi)(dr^2+r^2d\Omega^2).
\end{equation}
$\Psi$ and $\Phi$ correspond to the Newtonian gravitational potential and the perturbation to the spatial curvature, respectively.  Furthermore, due to the weak field limit, perturbations on time and space can be set as $\Phi=-\Psi$. The same relation is also valid for a cosmological scenario.  Since radiation strongly coupled to matter (coupling of the relativistic particles to the baryons) in the radiation dominated era and the energy density of radiation can be ignored in the matter dominated era,  radiation and matter behave like a fluid.  So, one can safely disregard the anisotropic stress in both epochs\cite{dodelson2003}. Thus, the perturbed FLRW metric takes the same form in the conformal Newtonian gauge, except the scale factor in front of its spatial part.

On the metric tensor in Eq.(10), the effect of gravitational potential term is quite apparent as perturbation. It alters time and space intervals for a space-time point by reshaping the metric components. These intervals are relative and dependent on the magnitude of the potential.  In the same manner, the total gravitational potential of the universe may affect the cosmological metric.  In the FLRW universe,  space is expanding physically with time and there is only one type of time passing with the same pace since the very beginning of the Big Bang. However, we live in a universe full of matter. So, in a way, the cosmological space-time intervals in the metric should be formed by a zeroth order potential term originating from the matter distribution. Since this potential term changes with time due to the expansion of space, our relative space and time intervals should also change according to any other observer standing on a different point on the time axis. So, there should be two kinds of time intervals. One is comoving and the other one is physical time just as for spatial intervals. One can perform a coordinate transformation to get back into the FLRW space-time in which a physical time coordinate is used, but a few arguments in the following section will be presented about why one should keep using the comoving one. In addition, if the matter distribution is assumed to be homogeneous at zeroth order, the gravitational potential and the related JBD scalar field should be independent of spatial coordinates.

To explain the relation in Eq.(8), the JBD theory proposes that the gravitational constant G is a time dependent parameter as mentioned in the introduction. There exist research papers\cite{uzan2011, dai2021, bhagvati2021, hofmann2018, williams2004, guenther1998,  bonanno2020,  corsico2013, verbiest2008, vijaykumar2021,ooba2016, ooba2017, wu2010, amirhashchi2020, alvey2020, chen2022} providing constraints on the possible variation of $G$ with time based on different observations.  It is also known from the Shapiro delay\cite{shapiro} that the speed of light seems to slow down due to the gravitational potential near an astrophysical object in the coordinate system of an observer that is infinitely far away from the object.  For a cosmological scenario in which the gravitational potential evolves with time, hypothesizing the speed of light time dependent would lead to inconsistencies.  However, it can safely be chosen as relatively changing among coordinate systems.  In this way, while the speed of light is a universal constant for all observers in their comoving coordinates, its value relatively changes (similar to the case of Shapiro delay) for observers separated by time. The future and the past light cones are different than those in the standard FLRW metric. This change in light cones results from relative cosmological space-time intervals. By taking the arguments and the observation mentioned above into consideration, we select $1/G=\phi\sim M/R$ as in the JBD theory, i.e. Eq.(9). Thus, the perturbed metric in Eq.(10) can be put in the following form by ignoring higher order terms

\begin{equation}
ds^2=-\left(1-\frac{2\delta\phi}{\phi_0}\right)dt^2+\left(1-\frac{2\delta\phi}{\phi_0}\right)^{-1}(dr^2+r^2d\Omega^2),
\end{equation}
where $\delta\phi$ represents contributions coming from inhomogeneities and $\phi_0$ is the present value of $\phi$ whose reciprocal stands for $G$. The metric form in Eq.(11) is valid for any region and can define a space-time point inside a non-expanding universe. The point of view that must be emphasized, however, is that any such $\delta\phi$ can be regarded as a perturbation normalized with respect to $\phi_0$ (i.e., $|\delta\phi|\ll \phi_0$) and it inversely affects time and space intervals.

When we look at any point in space, we are looking back in time that means we are getting photons from an era in which the value of $\phi$ is bigger than the value today due to the expansion of space. The difference in the field $\phi$ or the expanding space causes photons arriving us to be red-shifted.  For this reason, the time dependent scalar field must be in accord with the red-shift phenomenon i.e., the expansion of space. 

In addition, in Newtonian mechanics, the gravitational force is given by
\begin{equation}
\vec{F}=-m\vec{\nabla}\phi,
\end{equation}
so the field $\phi$ may take part in the components of a cosmological metric tensor to be able to define gravity in a purely geometrical way as done by Einstein himself for the general relativity. Therefore, in order to construct a metric for changing time and space intervals,  which are determined by the normalization of the scalar field with respect to its value today as in the case of Eq.(11), it is better to choose $\phi=\phi_0/a^2$ to embed the time dependence into the metric as the scale factor. Since the scale factor increases with time, $\phi$ will be decreasing correspondingly. Incidentally,  once the equation of $\phi$ and the solutions to the scale factor for different epochs are obtained in the next section,  it will be shown that our choice of $\phi=\phi_0/a^2$ is a solution. A zeroth order cosmological metric covering the properties above, after normalization with respect to $\phi_0$, should have the following form that can be called as the coordinate-transformed or shortly transformed FLRW metric.
\begin{equation}
ds^2=-\frac{dt^2}{a^2}+a^2\left(\frac{dr^2}{1-\kappa r^2}+r^2d\Omega^2\right)
\end{equation}

At this point, it may look like that $00$ component of the metric is the only change and a simple coordinate transformation compared to the FLRW metric. It is, however, more than that because the meaning behind the metric components has also been changed. There is no absolute time and space intervals for observers separated by time. They can be measured relatively because of the time dependence of the field $\phi$. Length and time intervals do not vary in a way that one is used to think for the FLRW metric. The contraction and the expansion of space-time intervals are what we are experiencing in our comoving coordinate system for the past and the future. Furthermore, whenever an observation takes place,  it should be considered as being performed from a new coordinate system because of change in $\phi$. This is a more appropriate and meaningful interpretation for a cosmological metric constructed based on Mach's principle. 

In the case of inhomogeneous space (Eq.(11)), light cones shrink a bit depending on the value of $\delta\phi$ in regions,  where $\delta\phi$ is bigger by comparison. In the cosmological case, a light cone at any time in the past, where $\phi$ is bigger, seems much more open in our comoving coordinate system. These two seem to contradict each other since we are considering a space-time point on which the fields are relatively bigger in both cases. However, while $\delta\phi$ changes with the spatial coordinate $r$ in the first case, it is the time coordinate assigning the coordinate system in the latter one.  Their opposite effects on the metric components can be explained by the scale-dependent gravity as in Mannheim's conformal gravity\cite{mannheim2000,mannheim2012}. Thus, gravity can be attractive at small scales but repulsive at cosmic scales and this can be the reason behind this apparent contradiction. In our model, it is the JBD parameter that will determine how gravity acts at different scales.

\section{Modified JBD Field Equation}
Since the FLRW metric has been coordinate-transformed, it is necessary to make all basic calculations for the new metric tensor. At this point it must be emphasized again that one can always make a coordinate transformation between comoving and physical times with the relation $dt/a(t)=dt^\prime$ to get back to the FLRW metric. Although solutions for both metric tensors will be mathematically equivalent, we continue with the transformed one (one written in comoving time) not to hide the effect of the field $\phi$ on time intervals. It is better to start with the action that is the same as that of the JBD theory.  However,  the metric tensor is now a function of the scalar field,  which is an important modification.
\begin{equation}
S=\frac{1}{16\pi}\int {d^4x\sqrt{-g}\left(\phi R-\omega g^{\mu\nu}(\phi)\frac{\partial_\mu\phi\partial_\nu\phi}{\phi}+16\pi\mathcal{L}_M(g_{\mu\nu}(\phi),\psi_i)\right)}
\end{equation}
Here,  the first term represents the non-minimal coupling of the scalar field to the Ricci scalar (we study in the Jordan frame) while the second one is the kinetic term of the field.  The third one is there because of the energy content of the universe coming from different matter fields $\psi_i$ as in any general relativistic Lagrangian density.  Since the matter Lagrangian is a function of the Jordan frame metric, test particles follow geodesics set by this metric tensor.

In the standard JBD theory,  there are two field equations obtained by varying the action with respect to $g_{\mu\nu}$ and $\phi$ (contains $\partial_\mu\phi$ variation).  In our case, however, there will be a single equation resulting from the variation with respect to $\phi$ since the metric tensor is also a function of $\phi$. Hence,  varying the action with respect to $\phi$ and $\partial_\mu\phi$ yields
\begin{equation}
\delta S=\frac{1}{16\pi}\int d^4x \left(\frac{\partial \mathcal{L}}{\partial g^{\mu\nu}}\frac{d g^{\mu\nu}}{d \phi}\delta \phi+\frac{\partial \mathcal{L}}{\partial \phi}\delta \phi+\frac{\partial \mathcal{L}}{\partial (\partial_\mu\phi)}\delta (\partial_\mu\phi)\right).
\end{equation}
The straightforward calculations can be carried out to obtain the field equation
\begin{equation}
\begin{split}
\left(R_{\mu\nu}-\frac{1}{2}R g_{\mu\nu}\right)\frac{d g^{\mu\nu}}{d \phi}= & \left[\frac{8\pi}{\phi}T_{\mu\nu}+\frac{1}{\phi}\left(\nabla_\mu\partial_\nu\phi-g_{\mu\nu}g^{\alpha\beta}\nabla_\alpha\partial_\beta\phi\right)\right.\\
& +\left.\frac{\omega}{\phi^2}\left(\partial_\mu\phi\partial_\nu\phi-\frac{1}{2}g_{\mu\nu}g^{\alpha\beta}\partial_\alpha\phi\partial_\beta\phi\right)\right]\frac{d g^{\mu\nu}}{d \phi}\\
&-\left(\frac{R}{\phi}+2\omega g^{\alpha\beta}\frac{\nabla_\alpha \partial_\beta \phi}{\phi^2}-\omega g^{\alpha\beta}\frac{\partial_\alpha \phi \partial_\beta \phi}{\phi^3}\right),
\end{split}
\end{equation}
where
\begin{equation}
T_{\mu\nu}=-\frac{2}{\sqrt{-g}}\frac{\delta{S_M}}{\delta{g^{\mu\nu}}}.
\end{equation}

In order to see how the matter and the radiation densities of the universe evolve with time,  one should look at the continuity equation.  Since test particles move on geodesics determined by the Jordan frame metric,  it is obtained by using the transformed FLRW metric for the relation $\nabla_\mu T^{\mu\nu}=0$ or simply by making a transformation for the time coordinate in Eq.(4).  This equation does not change under coordinate transformation and the continuity equation is still $\dot{\rho}+3\dot{a}/a(\rho+p)=0$.  So, the matter and the radiation densities evolve with the scale factor as in the case of the FLRW metric. Using $p_m=0$ for the pressure of ordinary matter (or dust) and $p_r=\rho_r/3$ for the pressure of radiation gives the evolution of the energy densities as
\begin{equation}
\rho_m=\frac{\rho_{m0}}{a^3} ,
\end{equation}
\begin{equation}
\rho_r=\frac{\rho_{r0}}{a^4} .
\end{equation}
From the extra $1/a$ factor in the evolution of the radiation density, it is easily seen that the wavelength of a photon is again proportional to the scale factor in the transformed FLRW metric. So, the relation in Eq.(6) is still valid.

As the next step,  the Ricci tensor components and the Ricci scalar for the transformed FLRW metric are calculated as
\begin{equation}
R_{00}=-3\left(\frac{\ddot{a}}{a}+\frac{\dot{a}^2}{a^2}\right) ,
\end{equation}
\begin{equation}
R_{11}=\frac{\ddot{a}a^3 + 3\dot{a}^2a^2 + 2\kappa}{1-\kappa r^2} ,
\end{equation}
\begin{equation}
R_{22}=r^2(\ddot{a}a^3 + 3\dot{a}^2a^2 + 2\kappa) ,
\end{equation}
\begin{equation}
R_{33}=r^2\sin^2\theta(\ddot{a}a^3 + 3\dot{a}^2a^2 + 2\kappa) ,
\end{equation}
\begin{equation}
R=6\left(\ddot{a}a + 2\dot{a}^2 + \frac{\kappa}{a^2}\right) .
\end{equation}
By using Eq.(20)-Eq.(24),  $\phi=\phi_0/a^2$ and the following relations for the summation in Eq.(16)
\begin{equation}
\frac{dg^{00}}{d \phi}=\frac{\phi_0}{\phi^2},
\end{equation}
\begin{equation}
\frac{dg^{11}}{d \phi}=\frac{1-\kappa r^2}{\phi_0},
\end{equation}
\begin{equation}
\frac{dg^{22}}{d \phi}=\frac{1}{\phi_0 r^2},
\end{equation}
\begin{equation}
\frac{dg^{33}}{d \phi}=\frac{1}{\phi_0 r^2sin^2\theta},
\end{equation}
the field equation is obtained as
\begin{equation}
\ddot{a}a=\frac{3}{3+2\omega}\left(\frac{4\pi}{3}\frac{(\rho+3p)}{\phi}-\frac{\kappa}{a^2}\right).
\end{equation}
By the definition
\begin{equation}
\omega=-\frac{3}{2}(1+\alpha),
\end{equation}
Eq.(29) reduces to
\begin{equation}
\ddot{a}a=-\frac{4\pi}{3\alpha\phi}(\rho+3p)+\frac{\kappa}{\alpha a^2}.
\end{equation}
As a field equation, we have just gained the acceleration equation, which is similar with the one in the standard model.  If one ignores the changes resulting from using the transformed FLRW metric, the differences are the curvature term and the factor of $\alpha$ in the denominators coming from the JBD parameter. Thus,  based on our model, the acceleration of the universe depends not only on ($\rho+3p$) but also on the spatial curvature, which is not the case in the standard model of cosmology.

In order to have the equation of $\dot{a}^2$, which is the other Friedmann-like equation,  one can firstly rewrite Eq.(31) by adding and subtracting $8\pi\rho/(3\alpha\phi)$ as
\begin{equation}
\ddot{a}a=-\frac{4\pi}{\alpha\phi}(\rho+p)+\frac{8\pi}{3\alpha\phi}\rho+\frac{\kappa}{\alpha a^2}
\end{equation}
and multiply Eq.(32) with $\dot{a}/a$ to get
\begin{equation}
\dot{a}\ddot{a}=-\frac{4\pi}{\alpha\phi}\frac{\dot{a}}{a}(\rho+p)+\frac{8\pi}{3\alpha\phi}\frac{\dot{a}}{a}\rho+\frac{\kappa\dot{a}}{\alpha a^3}.
\end{equation}
Then,  taking $\phi=\phi_0/a^2$ and using the continuity equation to get rid of pressure in Eq.(33) yields
\begin{equation}
\dot{a}\ddot{a}=\frac{4\pi}{3\alpha\phi_0}a^2\dot{\rho}+\frac{8\pi}{3\alpha\phi_0}a\dot{a}\rho+\frac{\kappa\dot{a}}{\alpha a^3}.
\end{equation}
Finally,  both sides of this equation can be written as time derivatives
\begin{equation}
\frac{1}{2}\frac{d}{dt}(\dot{a}^2)=\frac{d}{dt}\left(\frac{4\pi}{3\alpha\phi_0}a^2\rho-\frac{\kappa}{2\alpha a^2}\right)
\end{equation}
to obtain
\begin{equation}
\dot{a}^2=\bar{H}^2=\frac{8\pi}{3\alpha\phi}\rho-\frac{\kappa}{\alpha a^2}.
\end{equation}
Here,  as a new and more suitable parameter, the Hubble parameter is defined as $\bar{H}=\dot{a}$ since $v_{phy}=\dot{a}r_{phy}$ in our framework.  

At this point,  one can also look for the pressure equation by expanding Eq.(31) like
\begin{equation}
\ddot{a}a=-\frac{4\pi}{\alpha\phi}p-\left(\frac{4\pi}{3\alpha\phi}\rho-\frac{\kappa}{2\alpha a^2}\right)+\frac{\kappa}{2\alpha a^2},
\end{equation}
where the terms in the parenthesis is equal to $\dot{a}^2/2$.  Taking this term to the left hand side and multiplying whole equation with $-2$,  one ends up with the pressure equation
\begin{equation}
-2\ddot{a}a-\dot{a}^2=\frac{8\pi}{\alpha\phi}p-\frac{\kappa}{\alpha a^2}.
\end{equation}

From Eq.(36), one may conclude that the matter density dominates the evolution of the universe after the spatial curvature and the radiation density.  In addition, the scale factors for the matter, radiation and curvature dominated universe are found as $a\sim t^{2/3}$, $a\sim t^{1/2}$ and $a\sim t^{1/2}$ respectively. These power-law solutions may not seem interesting because they look like the same as in the standard model of cosmology,  except the one for the curvature dominated era.  However, it should be remembered that these solutions are obtained in our comoving time coordinate. In other words, while spatial intervals are expanding, time intervals will be getting smaller or contracted with the scale factor in the current scenario. Their combined effect can not be distinguished from the accelerating expansion of space when calculating the luminosity distance of a type Ia supernova. Instead, as it is stated before, one may get back to the FLRW metric via $dt^\prime=dt/a(t)$,  and obtain solutions such as $a\sim t^{\prime 2}$, $a\sim t^{\prime 1}$ and $a\sim t^{\prime 1}$ in physical time for the matter, radiation and curvature dominated universe in that order.  Now,  it is more obvious that the space undergoes a linear expansion in the era dominated by the curvature and the radiation, and an accelerating expansion in the matter dominated one.  However, we will stick to using comoving time or the former interpretation to avoid the impression that the ordinary matter or any other form of matter results in an accelerating expansion of space.  In addition to that, it would be a bit strange to use physical time while using a comoving space coordinate.  Eq.(36), which is a Friedmann-like equation, differs from the original one with the definition of the Hubble parameter and the time dependence of $1/G$,  which will be very helpful when solving the flatness and the horizon problems. 

When constructing the transformed FLRW metric,  on the arguments presented in the introduction, the evolution of the field has been chosen $1/a^2$. At this point, the time dependence of $\phi$ can be derived by taking an integral over all contributions coming from the observable universe. The quantity to be integrated is
\begin{equation}
d\phi=\frac{dM}{r_{ltr}},
\end{equation}
where $dM=4\pi r_{ltr}^2 \frac{\rho_0}{a^{3(1+w)}}dr_{ltr}$ and $r_{ltr}$ is the light travelled distance.  In comoving time, the distance taken by the gravitational effect of a shell of dust (it travels with the speed of light) for a general scale factor $a\sim t^s$ (since the solutions found in the previous section is power law) is
\begin{equation}
r_{ltr}=\int_{t^\prime}^{t} \frac{dt^{\prime\prime}}{a(t^{\prime\prime})}.
\end{equation}
Hence, the field at a point in space is calculated with the integral
\begin{equation}
\phi=4\pi \rho_0 \int_{0}^{t} \frac{dt^\prime}{a^{3(1+w)+1}(t^\prime)}\int_{t^\prime}^{t} \frac{dt^{\prime\prime}}{a(t^{\prime\prime})} .
\end{equation}
One can use $a\sim t^s$ and $w$ values for the matter and the radiation dominated epochs accordingly in Eq.(41) to find
\begin{equation}
\phi=\frac{4\pi \rho_0 t_0^2}{2s(1+s)}\left(\frac{t}{t_0}\right)^{-2s} .
\end{equation}
The result of the integration, Eq.(42), is valid for both eras, the type of energy content does not matter. Thus,  it confirms that the field is proportional to $1/a^2=(t/t_0)^{-2s}$, which agrees with our motivation. 

\section{Solutions to Flatness, Horizon and Late-time Accelerating Expansion Problems}
 
On the basis of the original Friedmann equation, the absolute value of the curvature density parameter increases with time in the matter and the radiation dominated eras. Since its value is nearly zero today or negligible\cite{2018planck, wilkinson} compared to other density parameters, it must have been extremely close to zero at very early times. In other words, the initial condition had to be fine tuned to $\kappa=0$, which is an unstable point. This is called the flatness problem in the standard model of cosmology. By using the solutions of the scale factor,  it can be checked that the following curvature density parameter defined based on Eq.(36) is constant in the radiation dominated era and decreases with time in the matter dominated era ($a^2\dot{a}^2$ is constant for radiation and $a^2\dot{a}^2\propto t^{2/3}$ for matter). 
\begin{equation}
|\Omega_\kappa|=|-\frac{\kappa}{\alpha a^2 \bar{H}^2}|=|-\frac{\kappa}{\alpha a^2\dot{a}^2}|
\end{equation}
According to Eq.(43), it does not matter which curvature parameter the universe had at the beginning. Even if it started to evolve as open or closed, the universe we see today is very likely to exist because the curvature density parameter is dragged to zero with time.  So, there is no fine-tuning or naturalness problem in this approach. This substantial change in the evolution of $|\Omega_\kappa|$ is actually a solution to the flatness problem.

Now,  we continue with the solution of the horizon problem, which is simply to understand how it is possible for the CMB to be nearly homogeneous for all parts of the sky although the comoving distance from decoupling (the period of time in which photons decoupled from combining protons and electrons and started to travel freely in space) to today is much bigger than the comoving particle horizon at the time of decoupling in the standard cosmology.  As a solution to this,  one can compare two comoving distances, which have been just mentioned,  in our approach.  In the transformed FLRW metric, a null path taken by a photon, which is the fastest way of interaction, is
\begin{equation}
\begin{split}
R_0 S_{\kappa}\left(R_0^{-1} \int \frac{dt}{a^2}\right) & =R_0 S_{\kappa}\left(R_0^{-1} \int \frac{da}{a^2 \bar{H}}\right)\\
&= R_0 S_{\kappa}\left((R_0 \bar{H}_0)^{-1} \int \frac{dz}{E(z)}\right),
\end{split}
\end{equation}
where $E(z)=\left[\Omega_{m 0}(1+z)+\Omega_{r 0}(1+z)^2+\Omega_{\kappa 0}(1+z)^2\right]^{1/2}$.  In order to calculate the comoving particle horizon at the period of decoupling one needs to figure out the redshift at which it occurred and so solve the Boltzmann equation.  Since our model does not make a notable change in the radiation and the matter densities (based on the fit with the Type Ia supernovae data which will be shown next),  the baryon to photon ratio is similar to the one in the conventional cosmology and the redshift of decoupling is predicted as $z_{*}\approx 1000$.  Considering the redshift of the Big Bang as infinite makes the integration in Eq.(44) to diverge.  Since our classical model should be valid at least up to the Planck scale,  the redshift at which the universe was as small as the Planck scale, can be taken as the upper limit of the integration.  By dividing the Planck temperature ($1.42\times 10^{32}$ K) by the temperature today ($2.73$ K), the redshift of the Planck scale is found to be $z_{Pl}\simeq 5\times 10^{31}$.  In addition,  for the energy content of the universe one can take $\Omega_{m0}\simeq 1$ and $\Omega_{r0}\simeq  10^{-4}$ as suggested by our model. Thus,  it is easy to show for a flat universe (for simplicity a flat universe scenario is taken into consideration but it can be shown for a closed and open universe as well) the comoving particle horizon at decoupling ($c=1$) is

\begin{equation}
\bar{H}_0^{-1} \int_{z_{*}}^{z_{Pl}} \frac{dz}{E(z)}\simeq 6454\ \bar{H}_0^{-1}
\end{equation}
and the comoving distance from decoupling to the present is
\begin{equation}
\bar{H}_0^{-1} \int_0^{z_{*}} \frac{dz}{E(z)}\simeq 60\ \bar{H}_0^{-1}.
\end{equation}
Eq.(45) and Eq.(46) show the observed comoving scale that is uniform in temperature is much smaller than the comoving particle horizon of decoupling.  Hence,  the horizon problem is eliminated.

In cosmology, providing the luminosity distance of a supernova in terms of the red-shift of photons in a way compatible with observations is an important theoretical competence that any successful theory should have. By using the ratio of flux to luminosity
\begin{equation}
\frac{F}{L}=\frac{1}{(1+z)^2 A} ,
\end{equation}
where $A=4\pi R_0^2 S_{\kappa}^2(\chi)$, $F$ is flux and $L$ is the actual luminosity of a standard candle, the relation for the luminosity distance is found as
\begin{equation}
d_L=(1+z) R_0 S_{\kappa}(\chi).
\end{equation}
In an expanding space, flux is reduced by a factor $(1+z)^2$ due to the red-shift of photons and the dilution of their number. This is why the factor $(1+z)^2$ lies in front of the comoving surface area $A$.  Although matter, radiation and curvature density parameters evolve as $\Omega_{m0}(1+z)^3$, $\Omega_{r0}(1+z)^4$ and $\Omega_{\kappa0}(1+z)^2$ in the right hand side of Eq.(36) respectively, their effective changes with the red-shift are $\Omega_{m0}(1+z)$,  $\Omega_{r0}(1+z)^2$ and $\Omega_{\kappa0}(1+z)^2$ since $\phi=\phi_0/a^2=\phi_0(1+z)^2$.  Thus, in our model,  a luminosity distance formula that may look similar to the one in Eq.(7) at the first sight is actually very different due to a totally different energy content and the factor of $\alpha$ coupled to the curvature density parameter
\begin{equation}
d_{L}(z)=(1+z) \frac{\bar{H}_{0}^{-1}}{\sqrt{\alpha |\Omega_{\kappa 0}|}}S_k\left(\sqrt{\alpha |\Omega_{\kappa 0}|} \int \frac{dz^{\prime}}{E(z^{\prime})}\right),
\end{equation}
where $E(z^{\prime})=\left[\Omega_{m 0}(1+z^{\prime})+\Omega_{r 0}(1+z^{\prime})^2+\Omega_{\kappa 0}(1+z^{\prime})^2\right]^{1/2}$. When the curvature and the radiation densities are neglected, Eq.(46) is simplified to
\begin{equation}
d_{L}(z)=(1+z) \bar{H}_{0}^{-1} \int (1+z^\prime)^{-1/2} dz^{\prime} .
\end{equation}

The data published by the Supernova Cosmology Project\cite{suzuki2012hubble} can be used to test the conformity between the observation and the theory for the late-time accelerating expansion.  Based upon our model, the expansion of space is not accelerating because the power of comoving time coordinate in the scale factor is smaller than one for all eras.  However, time intervals are contracted with time as well.  Hence,  the combined effect corresponds to an acceleration in the FLRW universe.  Since, in general, the redshift vs. distance modulus graph is more common among cosmologists, we prefer to use the following form of the distance modulus $\mu$
\begin{equation}
\mu=m-M=25+5 log_{10}\left(\frac{d_L}{Mpc}\right),
\end{equation}
where m and M are apparent and absolute magnitudes of an astronomical light source.

In Fig.(1), as dots represent the Type Ia supernovae data, the fitted solid curve is the relation between the redshift and the distance modulus for a flat universe predicted by our model (as a figure we put only the one for a flat case since the figures for closed and open cases look exactly the same).  The fit for a flat $\Lambda\mathrm{CDM}$ cosmology is shown with the dashed line.  Both are obtained with the weighted least square method.  Considering the best fit parameters, the flat universe is slightly preferred over closed and open universe in our model. In addition, the best fit values of the curvature density parameters in closed and open states are very close to zero.  Although we compare our model with the flat $\Lambda$CDM, which is currently accepted as a concordance model, there is an ongoing debate in the literature about the curvature of space\cite{di2020planck,handley2021curvature, efstathiou2020evidence, vagnozzi2021galaxy, vagnozzi2021eppur}. For a flat universe in our model,  the parameter $\alpha$ does not go into the fit and the Hubble constant that is the only free parameter,  is found $\bar{H}_0=70.37\ \mathrm{km/s/Mpc}$. The best-fit is achieved for a closed universe with $\Omega_{\kappa 0}=-1.21\times 10^{-7}$, $\alpha=0.55$ and $\bar{H}_0=70.37\ \mathrm{km/s/Mpc}$.  For an open universe, they are $\Omega_{\kappa 0}=9.04\times 10^{-7}$, $\alpha=0.18$ and $\bar{H}_0=70.37\ \mathrm{km/s/Mpc}$.  Note that the parameters in these fits are the ones that naturally arise in the JBD theory.  There is no need to add any energy source other than the matter (baryonic and dark) and the radiation.  Since,  $\phi_0$ does not go into this fit and the relation $\phi=1/G$ is our motivation from the beginning,  $\phi_0=1/G_0=1/(6.67\times 10^{-11} \ \mathrm{kg^{-1}\ m^3\ s^{-2}})\simeq 1.50\times 10^{10}\ \mathrm{kg^{-1}\ m^3\ s^{-2}}$,  where $G_0$ is the value of Newton's constant today.  By using fitted values and $\phi_0$, one can find today's values of the matter and the radiation densities as $\rho_{m0}\simeq 5.12\times 10^{-27}\  \mathrm{kg\ m^{-3}}$ and $\rho_{r0}\simeq 4.64\times 10^{-31}\  \mathrm{kg\ m^{-3}}$.  Although the value of $\alpha$ makes $\omega$ negative and the solar system observations predict it to be a large positive number, i.e. $\omega>10^4$, the JBD theory with a negative coupling parameter\cite{bertolami, senseshadri, sharif2012cosmic, banerjeepavon} is common in the literature and a viable scenario. 

In the flat $\Lambda\mathrm{CDM}$ scenario,  on the other hand,  the best fit parameters are  $\bar{H}_0=70.00\ \mathrm{km/s/Mpc}$ and $\Omega_{\Lambda 0}=0.72$. To compare the goodness of the fits of the two models,  we use $\chi^2$ statistics, Bayesian Information Criterion (BIC) and Akaike Information Criterion (AIC).  In our approach $\chi^2/dof$ values ($dof$ stands for the degree of freedom) are found as $\chi^2/dof=585.741/579=1.012$,  $\chi^2/dof=585.741/577=1.015$  and $\chi^2/dof=585.741/577=1.015$ for a flat,  closed and open universe respectively while in a flat $\Lambda\mathrm{CDM}$ it is $\chi^2/dof=562.227/578=0.973$. Using the $\chi^2$ values above and ignoring the constant term coming from the natural logarithm of the maximized value of the likelihood function in the BIC and AIC definitions,  they are,  for all cases,  calculated as follows.  For the flat case BIC=592 and AIC=588, for the closed and open case BIC=605 and AIC=592 whereas for the flat $\Lambda$CDM BIC=575 and AIC=566. Taking these values into consideration,  we can conclude that the flat $\Lambda\mathrm{CDM}$ cosmology is favored over our model.  However, the JBD scalar field decreasing as $1/a^2$ is able to elucidate not only the late-time accelerating expansion but also the flatness and the horizon problems.  From this point of view,  since our model does not need the dark energy as well as the inflationary era in the early universe,  it stands in a very different place from the $\Lambda$CDM cosmology and the extensions of the JBD theory.  Indeed,  incorporating the cosmological constant term $\Lambda$ in the JBD theory\cite{peracaula2019,peracaula2020} can provide a better fit and solve the Hubble tension\cite{pan2021} with a more slowly decreasing scalar field compared to the one evolving as $1/a^2$.  At this point,  however,  it is not necessary because we have not implemented our approach to the CMB and the LSS calculations yet (we are currently working on it).  In other words,  for now,  there is no tension for the Hubble constant in our model and we prefer to avoid adding any dark energy term.  

\begin{figure}
\includegraphics[scale=0.33]{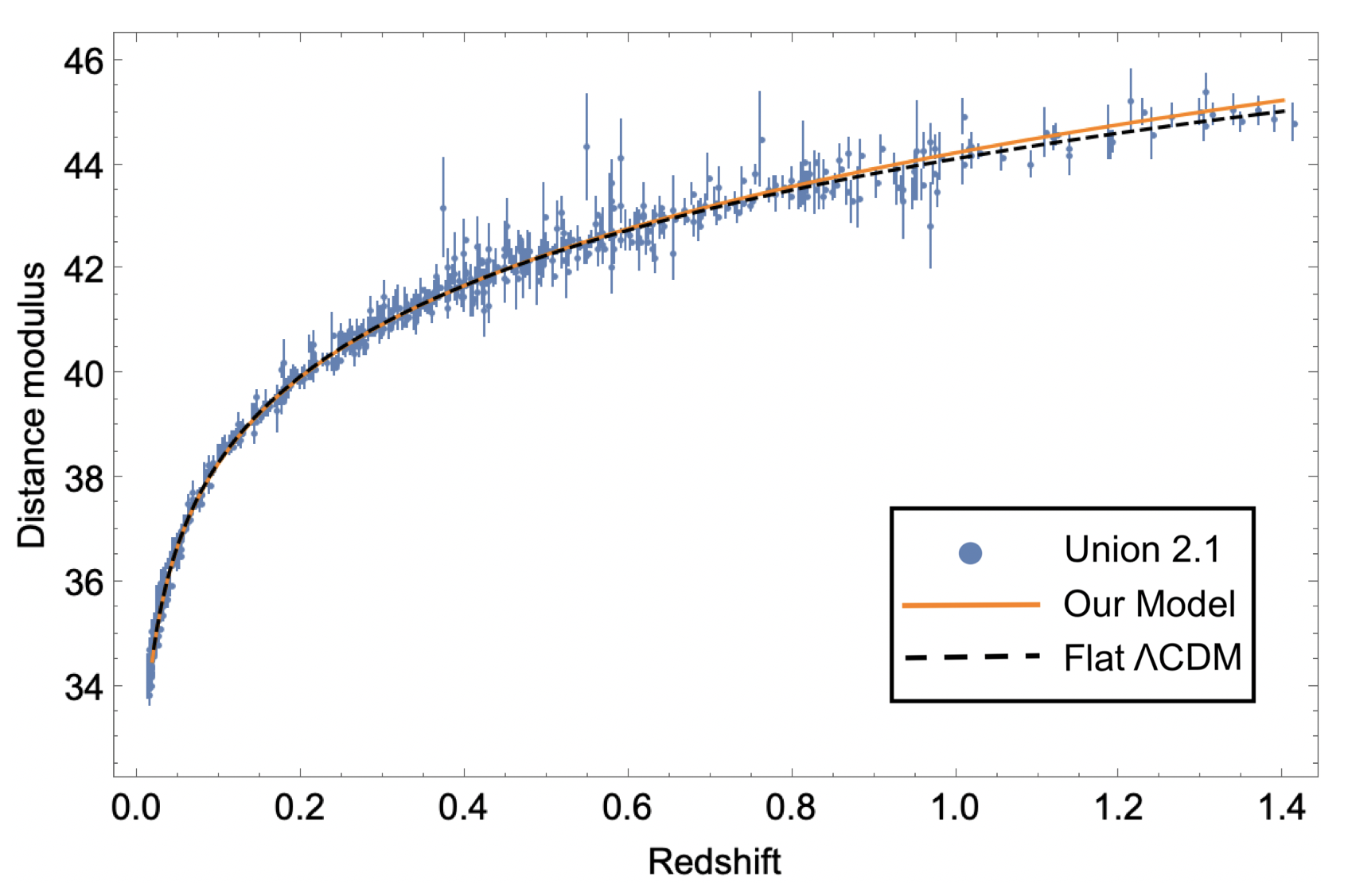}
\caption{ Redshift vs.  distance modulus diagram for the Union 2.1 Type Ia supernovae compilation. The solid orange line represents the best-fit for our model for a flat universe scenario and the dashed line is the best-fit for a flat $\Lambda\mathrm{ CDM}$ cosmology.  Whereas our approach has one free parameter ($\bar{H}$),  the flat $\Lambda\mathrm{CDM}$ has two ($\bar{H}$ and $\Omega_{\Lambda 0}$).}
\end{figure}

On the temporal evolution of $G$, there is a wide range of upper limits in the literature obtained from different observations.  Some of them are based on terrestrial measurements\cite{dai2021, bhagvati2021},  Lunar Laser Ranging\cite{hofmann2018, williams2004},  helioseismology\cite{guenther1998, bonanno2020}, the period change of pulsating white dwarfs\cite{corsico2013}, the precision timing of a millisecond pulsar\cite{verbiest2008} and the gravitational wave observation of a binary neutron star system\cite{vijaykumar2021}.  There are some others deduced from cosmological observations\cite{ooba2016, ooba2017, wu2010, amirhashchi2020, alvey2020, chen2022}, which are extremely dependent on the model chosen, such as the CMB, the Big Bang Nucleosynthesis (BBN) and the Baryon Acoustic Oscillations (BAO).  For a very detailed compilation, one may check \cite{uzan2011} by Uzan.  While the upper bounds on $|\dot{G}/G|_{t_0}$ from non-cosmological approaches range from $\sim 5\times\ 10^{-8}\ \mathrm{yr}^{-1}$ to $\sim 7\times\ 10^{-14}\ \mathrm{yr}^{-1}$,  the interval of cosmological ones is narrower and they vary between  $\sim 10^{-12}\ \mathrm{yr}^{-1}$ and $\sim 10^{-13}\ \mathrm{yr}^{-1}$.  The upper bounds in \cite{ooba2016, ooba2017} were given in terms of $G_{rec}/G_0$,  where $G_{rec}$ is the value of $G$ at recombination,  and they correspond to similar $|\dot{G}/G|_{t_0}$  bounds in the interval given above.  

In our case,  the variation of $G$ today can be found as
\begin{equation}
\frac{\dot{G}}{G}\bigg |_{t_0}=2\frac{\dot{a}}{a}\bigg |_{t_0}=2\bar{H}_0=1.44\times 10^{-10}\ \mathrm{yr}^{-1},
\end{equation}
which is not the stringent bound,  especially when the cosmological observations are taken into consideration.  However, the upper limits obtained from local and cosmological observations depend on the theory of gravity.  It should also be added that while obtaining such upper limits, in addition to choosing a gravity theory, some assumptions (related to the theory of gravity itself and/or other theoretical models in observation) are made within the theory (for details the review\cite{uzan2011} by Uzan can be checked).  For instance, the solar system constraints are sensitive to the models for asteroids.The same data but different asteroid models result in a difference of about one order of magnitude in $\dot{G}/G$.  Furthermore, they differ according to pulsar models in pulsar timing tests and nuclear reaction models for the interior of a star in surface temperature tests.  For the same pulsar,  the difference of about two orders of magnitude in $\dot{G}/G$ is obtained from different nuclear reaction models.  In cosmological observations,  it is necessary to consider how the temporal variation of $G$ affects the expansion of the universe from beginning to end.   So, using cosmological constraints derived from theories having different expansion histories to test a specific model of varying $G$ is not a healthy approach.   Even if the standard JBD theory was used to get these cosmological constraints,  one should not rely on them when making an inference about our model because we modified the JBD theory.  This modification brought many changes for the background equations regarding the spatial expansions in different epochs and the age of the universe,  both of which affect the BBN.  In addition to this,  the perturbation calculations for our approach are needed to check the consistency with the angular power spectrum of the CMB.  All of these will alter upper bounds on the variation of $G$ attained from cosmological observations.  At this point, it is very important to note that we do not question whether the mentioned limits or the way they are obtained are correct. We simply argue that these limits are highly dependent on the theoretical models (not only the theory of gravity but also other physical models in the observations).  Therefore,  theories with varying Newton's constant not obeying the strictest upper bounds derived based on another model may still be valid.

The age of the universe is another parameter that gives some constraints for a theoretical model and is calculated by measuring the Hubble constant. One can make an estimation about its value just by regarding the evolution of matter, radiation and curvature.  So, the age of the universe is given by
\begin{equation}
t_0=\int_0^{t_0} dt=\int_0^1 \frac{da}{\dot{a}}=\int_0^1\frac{da}{\bar{H}_0\sqrt{(1+a_{eq}a^{-1})\Omega_{m0} a^{-1}+\Omega_{\kappa0}a^{-2}}},
\end{equation}
where $dt=da/\dot{a}$ is used to change the variable,  and for $\dot{a}$ we have made use of Eq.(36).  As mentioned before,  although matter, radiation and curvature density parameters evolve as $\Omega_{m0}/a^3$, $\Omega_{r0}/a^4$ and $\Omega_{\kappa0}/a^2$ in Eq.(36) respectively, their effective changes with the scale factor are $\Omega_{m0}/a$,  $\Omega_{r0}/a^2$ and $\Omega_{\kappa0}/a^2$ due to $\phi=\phi_0/a^2$. In addition, $a_{eq}$ is the value of the scale factor at which the matter and the radiation densities are equal (so $\Omega_{r0}=a_{eq}\Omega_{m0}$).  The value calculated from Eq.(53) for the fitted parameters is smaller than the age calculated in the $\Lambda$CDM cosmology. So, our model seems to contradict some astrophysical constraints\cite{ageconstraint} related to the age of oldest stars. However, Eq.(53) gives the amount of time elapsed in comoving time.  The physical age, which is the actual age, is calculated as
\begin{equation}
\int_0^{t_0}\frac{dt}{a}=\int_0^1\frac{da}{a{\dot{a}}}=\int_0^1\frac{da}{\bar{H}_0\sqrt{(1+a_{eq}a^{-1})\Omega_{m0}a+\Omega_{\kappa0}}}.
\end{equation}
Now, for the best-fit parameters, one can find $27.6\times 10^9$ years, which is a bigger value in comparison with the age in the comoving time.  So, even the oldest stars had enough time to be in their present phase in our model.

\section{Conclusions}

In order to build the approach presented in this manuscript, one of the main motivations is the relation indicated in Eq.(8). It is a simple but an important relation because of the connection between the gravitational potential and the scalar field $\phi=1/G$ in the JBD theory. More significantly, however, it is the idea that the scalar field could be related to the components of the cosmological metric as it does as a potential term in other circumstances in the general theory of relativity.  Since there is a time dependent gravitational potential of the matter distribution related to the scalar field,  it seems obvious that our space-time intervals must be relative among coordinate systems when this field changes with time.  So, based on this, we have made use of the connection in Eq.(8) to select the JBD scalar field proportional to $1/a^2$. Then, the FLRW metric has been coordinate transformed to embed the concept of relative space-time in the new metric tensor. Once the metric is constructed, it is straightforward to do all basic calculations like finding the Ricci tensor components, the Ricci scalar and the field equation. Solutions for the flatness and the horizon problems have naturally come into the picture without any further modification apart from the one mentioned above. Lastly, it has been shown that the luminosity distance vs. red-shift data of Type Ia supernovae is in agreement with the prediction of our approach. The best-fit values for the Hubble constant for all spatial scenarios are the same and $\bar{H}_0=70.37$ km/s/Mpc.  The curvature density parameter and the JBD constant are found as $\Omega_{\kappa 0}=-1.21\times 10^{-7}$ and $\omega=-3/2(1+0.55)\approx-2.32$ for a closed universe.  For an open universe they are $\Omega_{\kappa 0}=9.04\times 10^{-7}$ and $\omega=-3/2(1+0.18)\approx-1.77$.  Although the scale factor for the matter dominated universe is still given by $a\sim t^{2/3}$ and there is no accelerating expansion in our comoving time, the combined effect of the contraction in time  and the expansion in space intervals makes supernovae to be observed dimmer in our comoving coordinates. Among them, the last one is the most important success of this new approach in comparison with the standard model. There may no need for the cosmological constant or other forms of dark energy in this model to conform with the supernovae data. Embedding the time dependent scalar field into the JBD theory in a particular way can solve three major problems of the standard model of cosmology.

\nocite{*}
\bibliographystyle{ieeetr}
\bibliography{references}

\begin{thebibliography}{10}

\bibitem{2018planck}
N.~Aghanim, Y.~Akrami, M.~Ashdown, J.~Aumont, C.~Baccigalupi, M.~Ballardini,
  A.~Banday, R.~Barreiro, N.~Bartolo, S.~Basak, {\em et~al.}, ``Planck 2018
  results-vi. cosmological parameters,'' {\em Astronomy \& Astrophysics},
  vol.~641, p.~A6, 2020.

\bibitem{carroll}
S.~M. Carroll, {\em Spacetime and geometry}.
\newblock Cambridge University Press, 2019.

\bibitem{1999perlmutter}
S.~Perlmutter, G.~Aldering, G.~Goldhaber, R.~Knop, P.~Nugent, P.~Castro,
  S.~Deustua, S.~Fabbro, A.~Goobar, D.~Groom, {\em et~al.}, ``Measurements of
  $\omega$ and $\lambda$ from 42 high-redshift supernovae,'' {\em The
  Astrophysical Journal}, vol.~517, no.~2, p.~565, 1999.

\bibitem{1998riess}
A.~G. Riess, A.~V. Filippenko, P.~Challis, A.~Clocchiatti, A.~Diercks, P.~M.
  Garnavich, R.~L. Gilliland, C.~J. Hogan, S.~Jha, R.~P. Kirshner, {\em
  et~al.}, ``Observational evidence from supernovae for an accelerating
  universe and a cosmological constant,'' {\em The Astronomical Journal},
  vol.~116, no.~3, p.~1009, 1998.

\bibitem{cosmiccoincidence}
I.~Zlatev, L.~Wang, and P.~J. Steinhardt, ``Quintessence, cosmic coincidence,
  and the cosmological constant,'' {\em Physical Review Letters}, vol.~82,
  no.~5, p.~896, 1999.

\bibitem{weinberg-cosmologicalconstant}
S.~Weinberg, ``The cosmological constant problem,'' {\em Reviews of modern
  physics}, vol.~61, no.~1, p.~1, 1989.

\bibitem{martin2012}
J.~Martin, ``Everything you always wanted to know about the cosmological
  constant problem (but were afraid to ask),'' {\em Comptes Rendus Physique},
  vol.~13, no.~6-7, pp.~566--665, 2012.

\bibitem{nobbenhuis}
S.~Nobbenhuis, ``Categorizing different approaches to the cosmological constant
  problem,'' {\em Foundations of Physics}, vol.~36, no.~5, pp.~613--680, 2006.

\bibitem{joyce2016}
A.~Joyce, L.~Lombriser, and F.~Schmidt, ``Dark energy versus modified
  gravity,'' {\em Annual Review of Nuclear and Particle Science}, vol.~66,
  pp.~95--122, 2016.

\bibitem{jordan}
P.~Jordan, ``Zum gegenw{\"a}rtigen stand der diracschen kosmologischen
  hypothesen,'' {\em Zeitschrift f{\"u}r Physik}, vol.~157, no.~1,
  pp.~112--121, 1959.

\bibitem{bransdicke}
C.~Brans and R.~H. Dicke, ``Mach's principle and a relativistic theory of
  gravitation,'' {\em Physical review}, vol.~124, no.~3, p.~925, 1961.

\bibitem{bertolami}
O.~Bertolami and P.~Martins, ``Nonminimal coupling and quintessence,'' {\em
  Physical Review D}, vol.~61, no.~6, p.~064007, 2000.

\bibitem{sen2001late}
S.~Sen and A.~Sen, ``Late time acceleration in brans-dicke cosmology,'' {\em
  Physical Review D}, vol.~63, no.~12, p.~124006, 2001.

\bibitem{senseshadri}
S.~Sen and T.~Seshadri, ``Self interacting brans--dicke cosmology and
  quintessence,'' {\em International Journal of Modern Physics D}, vol.~12,
  no.~03, pp.~445--460, 2003.

\bibitem{sharif2012cosmic}
M.~Sharif and S.~Waheed, ``Cosmic evolution in self-interacting brans--dicke
  cosmology,'' {\em Journal of the Physical Society of Japan}, vol.~81, no.~11,
  p.~114901, 2012.

\bibitem{bisabr2012cosmic}
Y.~Bisabr, ``Cosmic acceleration in brans--dicke cosmology,'' {\em General
  Relativity and Gravitation}, vol.~44, pp.~427--435, 2012.

\bibitem{chakrabarti2022screening}
S.~Chakrabarti, K.~Dutta, and J.~L. Said, ``Screening mechanism and late-time
  cosmology: Role of a chameleon--brans--dicke scalar field,'' {\em Monthly
  Notices of the Royal Astronomical Society}, vol.~514, no.~1, pp.~427--439,
  2022.

\bibitem{guth}
A.~H. Guth, ``Inflationary universe: A possible solution to the horizon and
  flatness problems,'' {\em Physical Review D}, vol.~23, no.~2, p.~347, 1981.

\bibitem{linde}
A.~D. Linde, ``A new inflationary universe scenario: a possible solution of the
  horizon, flatness, homogeneity, isotropy and primordial monopole problems,''
  {\em Physics Letters B}, vol.~108, no.~6, pp.~389--393, 1982.

\bibitem{turok2002critical}
N.~Turok, ``A critical review of inflation,'' {\em Classical and Quantum
  Gravity}, vol.~19, no.~13, p.~3449, 2002.

\bibitem{brandenberger2000inflationary}
R.~H. Brandenberger, ``Inflationary cosmology: Progress and problems,'' {\em
  ASTROPHYSICS AND SPACE SCIENCE LIBRARY}, vol.~247, pp.~169--212, 2000.

\bibitem{martin2001trans}
J.~Martin and R.~H. Brandenberger, ``Trans-planckian problem of inflationary
  cosmology,'' {\em Physical Review D}, vol.~63, no.~12, p.~123501, 2001.

\bibitem{brandenberger2007conceptual}
R.~H. Brandenberger, ``Conceptual problems of inflationary cosmology and a new
  approach to cosmological structure formation,'' {\em Inflationary Cosmology},
  pp.~393--424, 2007.

\bibitem{brandenberger2013trans}
R.~H. Brandenberger and J.~Martin, ``Trans-planckian issues for inflationary
  cosmology,'' {\em Classical and Quantum Gravity}, vol.~30, no.~11, p.~113001,
  2013.

\bibitem{la1989extended}
D.~La and P.~J. Steinhardt, ``Extended inflationary cosmology,'' {\em Physical
  Review Letters}, vol.~62, no.~4, p.~376, 1989.

\bibitem{mathiazhagan1984inflationary}
C.~Mathiazhagan and V.~Johri, ``An inflationary universe in brans-dicke theory:
  a hopeful sign of theoretical estimation of the gravitational constant,''
  {\em Classical and Quantum Gravity}, vol.~1, no.~2, p.~L29, 1984.

\bibitem{weinberg1989some}
E.~J. Weinberg, ``Some problems with extended inflation,'' {\em Physical Review
  D}, vol.~40, no.~12, p.~3950, 1989.

\bibitem{la1989prescription}
D.~La, P.~J. Steinhardt, and E.~W. Bertschinger, ``Prescription for successful
  extended inflation,'' {\em Physics Letters B}, vol.~231, no.~3, pp.~231--236,
  1989.

\bibitem{linde1994hybrid}
A.~Linde, ``Hybrid inflation,'' {\em Physical Review D}, vol.~49, no.~2,
  p.~748, 1994.

\bibitem{tahmasebzadeh2016brans}
B.~Tahmasebzadeh, K.~Rezazadeh, and K.~Karami, ``Brans-dicke inflation in light
  of the planck 2015 data,'' {\em Journal of Cosmology and Astroparticle
  Physics}, vol.~2016, no.~07, p.~006, 2016.

\bibitem{baptista1996density}
J.~Baptista, J.~Fabris, and S.~Gon{\c{c}}alves, ``Density perturbations in the
  brans-dicke theory,'' {\em Astrophysics and Space Science}, vol.~246,
  pp.~315--331, 1996.

\bibitem{kolb1990origin}
E.~W. Kolb, D.~S. Salopek, and M.~S. Turner, ``Origin of density fluctuations
  in extended inflation,'' {\em Physical Review D}, vol.~42, no.~12, p.~3925,
  1990.

\bibitem{acquaviva2007observational}
V.~Acquaviva and L.~Verde, ``Observational signatures of jordan--brans--dicke
  theories of gravity,'' {\em Journal of Cosmology and Astroparticle Physics},
  vol.~2007, no.~12, p.~001, 2007.

\bibitem{acquaviva2005structure}
V.~Acquaviva, C.~Baccigalupi, S.~M. Leach, A.~R. Liddle, and F.~Perrotta,
  ``Structure formation constraints on the jordan-brans-dicke theory,'' {\em
  Physical Review D}, vol.~71, no.~10, p.~104025, 2005.

\bibitem{chen1999cosmic}
X.~Chen and M.~Kamionkowski, ``Cosmic microwave background temperature and
  polarization anisotropy in brans-dicke cosmology,'' {\em Physical Review D},
  vol.~60, no.~10, p.~104036, 1999.

\bibitem{wu2010cosmic}
F.-Q. Wu and X.~Chen, ``Cosmic microwave background with brans-dicke gravity.
  ii. constraints with the wmap and sdss data,'' {\em Physical Review D},
  vol.~82, no.~8, p.~083003, 2010.

\bibitem{li2013constraints}
Y.-C. Li, F.-Q. Wu, and X.~Chen, ``Constraints on the brans-dicke gravity
  theory with the planck data,'' {\em Physical Review D}, vol.~88, no.~8,
  p.~084053, 2013.

\bibitem{ballardini2019testing}
M.~Ballardini, D.~Sapone, C.~Umilt{\`a}, F.~Finelli, and D.~Paoletti, ``Testing
  extended jordan-brans-dicke theories with future cosmological observations,''
  {\em Journal of Cosmology and Astroparticle Physics}, vol.~2019, no.~05,
  p.~049, 2019.

\bibitem{joudaki2022testing}
S.~Joudaki, P.~G. Ferreira, N.~A. Lima, and H.~A. Winther, ``Testing gravity on
  cosmic scales: A case study of jordan-brans-dicke theory,'' {\em Physical
  Review D}, vol.~105, no.~4, p.~043522, 2022.

\bibitem{dodelson2003}
S.~Dodelson, {\em Modern cosmology}.
\newblock Elsevier, 2003.

\bibitem{uzan2011}
J.-P. Uzan, ``Varying constants, gravitation and cosmology,'' {\em Living
  reviews in relativity}, vol.~14, no.~1, pp.~1--155, 2011.

\bibitem{dai2021}
D.-C. Dai, ``Variance of newtonian constant from local gravitational
  acceleration measurements,'' {\em Physical Review D}, vol.~103, no.~6,
  p.~064059, 2021.

\bibitem{bhagvati2021}
S.~Bhagvati and S.~Desai, ``Search for variability in newton’s constant using
  local gravitational acceleration measurements,'' {\em Classical and Quantum
  Gravity}, vol.~39, no.~1, p.~017001, 2021.

\bibitem{hofmann2018}
F.~Hofmann and J.~M{\"u}ller, ``Relativistic tests with lunar laser ranging,''
  {\em Classical and Quantum Gravity}, vol.~35, no.~3, p.~035015, 2018.

\bibitem{williams2004}
J.~G. Williams, S.~G. Turyshev, and D.~H. Boggs, ``Progress in lunar laser
  ranging tests of relativistic gravity,'' {\em Physical Review Letters},
  vol.~93, no.~26, p.~261101, 2004.

\bibitem{guenther1998}
D.~B. Guenther, L.~Krauss, and P.~Demarque, ``Testing the constancy of the
  gravitational constant using helioseismology,'' {\em The Astrophysical
  Journal}, vol.~498, no.~2, p.~871, 1998.

\bibitem{bonanno2020}
A.~Bonanno and H.-E. Fr{\"o}hlich, ``A new helioseismic constraint on a
  cosmic-time variation of g,'' {\em The Astrophysical Journal Letters},
  vol.~893, no.~2, p.~L35, 2020.

\bibitem{corsico2013}
A.~H. C{\'o}rsico, L.~G. Althaus, E.~Garc{\'\i}a-Berro, and A.~D. Romero, ``An
  independent constraint on the secular rate of variation of the gravitational
  constant from pulsating white dwarfs,'' {\em Journal of Cosmology and
  Astroparticle Physics}, vol.~2013, no.~06, p.~032, 2013.

\bibitem{verbiest2008}
J.~P. Verbiest, M.~Bailes, W.~van Straten, G.~B. Hobbs, R.~T. Edwards, R.~N.
  Manchester, N.~Bhat, J.~M. Sarkissian, B.~A. Jacoby, and S.~R. Kulkarni,
  ``Precision timing of psr j0437--4715: an accurate pulsar distance, a high
  pulsar mass, and a limit on the variation of newton’s gravitational
  constant,'' {\em The Astrophysical Journal}, vol.~679, no.~1, p.~675, 2008.

\bibitem{vijaykumar2021}
A.~Vijaykumar, S.~J. Kapadia, and P.~Ajith, ``Constraints on the time variation
  of the gravitational constant using gravitational wave observations of binary
  neutron stars,'' {\em Physical Review Letters}, vol.~126, no.~14, p.~141104,
  2021.

\bibitem{ooba2016}
J.~Ooba, K.~Ichiki, T.~Chiba, and N.~Sugiyama, ``Planck constraints on
  scalar-tensor cosmology and the variation of the gravitational constant,''
  {\em Physical Review D}, vol.~93, no.~12, p.~122002, 2016.

\bibitem{ooba2017}
J.~Ooba, K.~Ichiki, T.~Chiba, and N.~Sugiyama, ``Cosmological constraints on
  scalar--tensor gravity and the variation of the gravitational constant,''
  {\em Progress of Theoretical and Experimental Physics}, vol.~2017, no.~4,
  p.~043E03, 2017.

\bibitem{wu2010}
F.-Q. Wu and X.~Chen, ``Cosmic microwave background with brans-dicke gravity.
  ii. constraints with the wmap and sdss data,'' {\em Physical Review D},
  vol.~82, no.~8, p.~083003, 2010.

\bibitem{amirhashchi2020}
H.~Amirhashchi and A.~K. Yadav, ``Constraining an exact brans--dicke gravity
  theory with recent observations,'' {\em Physics of the Dark Universe},
  vol.~30, p.~100711, 2020.

\bibitem{alvey2020}
J.~Alvey, N.~Sabti, M.~Escudero, and M.~Fairbairn, ``Improved bbn constraints
  on the variation of the gravitational constant,'' {\em The European Physical
  Journal C}, vol.~80, no.~2, pp.~1--6, 2020.

\bibitem{chen2022}
A.~Chen, Y.~Gong, F.~Wu, Y.~Wang, and X.~Chen, ``Constraining brans--dicke
  cosmology with the csst galaxy clustering spectroscopic survey,'' {\em
  Research in Astronomy and Astrophysics}, vol.~22, no.~5, p.~055021, 2022.

\bibitem{shapiro}
I.~I. Shapiro, ``Fourth test of general relativity,'' {\em Physical Review
  Letters}, vol.~13, no.~26, p.~789, 1964.

\bibitem{mannheim2000}
P.~D. Mannheim, ``Attractive and repulsive gravity,'' {\em Foundations of
  Physics}, vol.~30, no.~5, pp.~709--746, 2000.

\bibitem{mannheim2012}
P.~D. Mannheim, ``Making the case for conformal gravity,'' {\em Foundations of
  Physics}, vol.~42, no.~3, pp.~388--420, 2012.

\bibitem{wilkinson}
J.~Dunkley, E.~Komatsu, M.~Nolta, D.~Spergel, D.~Larson, G.~Hinshaw, L.~Page,
  C.~Bennett, B.~Gold, N.~Jarosik, {\em et~al.}, ``Five-year wilkinson
  microwave anisotropy probe* observations: Likelihoods and parameters from the
  wmap data,'' {\em The Astrophysical Journal Supplement Series}, vol.~180,
  no.~2, p.~306, 2009.

\bibitem{suzuki2012hubble}
N.~Suzuki, D.~Rubin, C.~Lidman, G.~Aldering, R.~Amanullah, K.~Barbary,
  L.~Barrientos, J.~Botyanszki, M.~Brodwin, N.~Connolly, {\em et~al.}, ``The
  hubble space telescope cluster supernova survey. v. improving the dark-energy
  constraints above z> 1 and building an early-type-hosted supernova sample,''
  {\em The Astrophysical Journal}, vol.~746, no.~1, p.~85, 2012.

\bibitem{di2020planck}
E.~Di~Valentino, A.~Melchiorri, and J.~Silk, ``Planck evidence for a closed
  universe and a possible crisis for cosmology,'' {\em Nature Astronomy},
  vol.~4, no.~2, pp.~196--203, 2020.

\bibitem{handley2021curvature}
W.~Handley, ``Curvature tension: evidence for a closed universe,'' {\em
  Physical Review D}, vol.~103, no.~4, p.~L041301, 2021.

\bibitem{efstathiou2020evidence}
G.~Efstathiou and S.~Gratton, ``The evidence for a spatially flat universe,''
  {\em Monthly Notices of the Royal Astronomical Society: Letters}, vol.~496,
  no.~1, pp.~L91--L95, 2020.

\bibitem{vagnozzi2021galaxy}
S.~Vagnozzi, E.~Di~Valentino, S.~Gariazzo, A.~Melchiorri, O.~Mena, and J.~Silk,
  ``The galaxy power spectrum take on spatial curvature and cosmic
  concordance,'' {\em Physics of the Dark Universe}, vol.~33, p.~100851, 2021.

\bibitem{vagnozzi2021eppur}
S.~Vagnozzi, A.~Loeb, and M.~Moresco, ``Eppur {\`e} piatto? the cosmic
  chronometers take on spatial curvature and cosmic concordance,'' {\em The
  Astrophysical Journal}, vol.~908, no.~1, p.~84, 2021.

\bibitem{banerjeepavon}
N.~Banerjee and D.~Pavon, ``Cosmic acceleration without quintessence,'' {\em
  Physical Review D}, vol.~63, no.~4, p.~043504, 2001.

\bibitem{peracaula2019}
J.~S. Peracaula, A.~G{\'o}mez-Valent, J.~de~Cruz~P{\'e}rez, and
  C.~Moreno-Pulido, ``Brans--dicke gravity with a cosmological constant
  smoothes out $\lambda$cdm tensions,'' {\em The Astrophysical Journal
  Letters}, vol.~886, no.~1, p.~L6, 2019.

\bibitem{peracaula2020}
J.~S. Peracaula, A.~G{\'o}mez-Valent, J.~de~Cruz~P{\'e}rez, and
  C.~Moreno-Pulido, ``Brans--dicke cosmology with a $\lambda$-term: a possible
  solution to $\lambda$cdm tensions,'' {\em Classical and Quantum Gravity},
  vol.~37, no.~24, p.~245003, 2020.

\bibitem{pan2021}
S.~Pan, L.~Visinelli, W.~Yang, A.~Melchiorri, D.~F. Mota, A.~G. Riess, J.~Silk,
  {\em et~al.}, ``In the realm of the hubble tension-a review of solutions,''
  {\em Classical and Quantum Gravity}, vol.~38, no.~15, p.~153001, 2021.

\bibitem{ageconstraint}
L.~M. Krauss and B.~Chaboyer, ``Age estimates of globular clusters in the milky
  way: constraints on cosmology,'' {\em Science}, vol.~299, no.~5603,
  pp.~65--69, 2003.

\end{thebibliography}

\end{document}